\documentclass[fleqn,usenatbib]{mnras}

\usepackage[T1]{fontenc}
\usepackage{ae,aecompl}
\usepackage{CJK}
\usepackage{upgreek}
\usepackage{booktabs}
\usepackage[utf8]{inputenc}

\usepackage{graphicx}	
\usepackage{amsmath}	
\usepackage{amssymb}	

\usepackage{booktabs}
\usepackage{afterpage}
\usepackage{emptypage}
\usepackage{pdflscape}
\usepackage{cleveref}
\usepackage{multirow}
\usepackage{rotating}
\usepackage{verbatim}
\usepackage{threeparttable}
\usepackage[caption = false]{subfig}
\usepackage[colorinlistoftodos]{todonotes}
\usepackage{newtxtext,newtxmath}

\newcommand{\D} {\mathrm{d}}
\newcommand{\EQ}[1] {Equation~(\ref{#1})}
\newcommand{\FIG}[1] {Figure~\ref{#1}}

\newcommand{\SEC}[1] {Section~\ref{#1}}
\newcommand{\APP}[1] {Appendix~\ref{#1}}

\def\Sig{\mathcal{S}}
\def\Noi{\mathcal{N}}
\def\DM{\mathrm{DM}}

\def\cmpc{\mathrm{cm}^{-3}\, \mathrm{pc}}

\title[Simulating high-time resolution observations]{Simulating high-time resolution radio-telescope observations}
\author[R. Luo et al.]{
Rui Luo,$^{1}$\thanks{E-mail: rui.luo@csiro.au (RL)}
George Hobbs,$^{1}$\thanks{E-mail: george.hobbs@csiro.au (GH)}
Suk Yee Yong,$^{1}$
Andrew Zic,$^{2,1}$
Lawrence Toomey,$^{1}$
Shi Dai,$^{3}$
\newauthor
Alex Dunning,$^{1}$
Di Li,$^{4,5}$
Tommy Marshman,$^{2,1}$
Chen Wang,$^{6}$
Pei Wang,$^{4}$
Shuangqiang Wang$^{7}$
\newauthor
and Songbo Zhang$^{8}$
\\
\\
$^{1}$CSIRO Space and Astronomy, Australia Telescope National Facility, PO Box 76, Epping, NSW 1710, Australia\\
$^{2}$Department of Physics and Astronomy, and Research Centre in Astronomy, Astrophysics and Astrophotonics, Macquarie University, NSW 2109, Australia \\
$^{3}$School of Science, Western Sydney University, Locked Bag 1797, Penrith South DC, NSW 2751, Australia \\
$^{4}$National Astronomical Observatories, Chinese Academy of Sciences, A20 Datun Road, Chaoyang District, Beijing 100101, China \\
$^{5}$University of Chinese Academy of Sciences, Beijing 100049, China \\
$^{6}$CSIRO Data61, 13 Garden Street, Eveleigh, NSW 2015, Australia \\
$^{7}$Xinjiang Astronomical Observatory, Chinese Academy of Sciences, 150 Science 1-Street, Urumqi 830011, China \\
$^{8}$Purple Mountain Observatory, Chinese Academy of Sciences, Nanjing 210008, China \\
}

\date{Accepted XXX. Received YYY; in original form ZZZ}

\pubyear{2022}

\begin{document}
\label{firstpage}
\pagerange{\pageref{firstpage}--\pageref{lastpage}}
\maketitle

\begin{abstract}
We describe a new software package for simulating channelised, high-time resolution data streams from radio telescopes. The software simulates data from the telescope and observing system taking into account the observation strategy, receiver system and digitisation.  The signatures of pulsars, fast radio bursts and flare stars are modelled, including frequency-dependent effects such as scattering and scintillation.  We also simulate more generic signals using spline curves and images. Models of radio frequency interference include signals from satellites, terrestrial transmitters and impulsive, broadband signals. The simulated signals can also be injected into real data sets. Uses of this software include the production of machine learning training data sets, development and testing of new algorithms to search for anomalous patterns and to characterise processing pipelines.
\end{abstract}




\section{Introduction}

Searches that are carried out with radio telescopes for astronomical sources can often be divided into high-time and high-frequency resolution surveys. For the former the data streams are channelised with relatively poor frequency resolution (typically megahertz channel widths), but sampled with microsecond to millisecond time resolution.  Such surveys have discovered the majority of  known pulsars \citep{Hewish+68Nat, Manchester+01MN} and fast radio bursts (FRBs, \citealt{Lorimer+07Sci}). High-frequency resolution surveys typically have much higher frequency resolution (kHz channel bandwidths are common), but usually only record a sample every second, or so.  

High-time resolution surveys are ongoing at many observatories and new surveys are planned with the most sensitive radio telescopes. Examples include the Commensal Radio Astronomy FAST survey (CRAFTS, \citealt{Li+18IMMag}) carried by the Five-hundred-meter Aperture Spherical radio Telescope (FAST) and, within a year, surveys will begin with a new cryogenically-cooled phased-array receiver at the Parkes ``Murriyang'' radio telescope (Dunning et al., in prep.). In addition, wide-field, beam-formed observations are being designed for the TRAnsients and PUlsars with MeerKAT (TRAPUM) survey \citep{Chen+21JAI}. 

Processing and archiving data from high-time resolution surveys are challenging because of the massive data volumes.  The processing algorithms now must also deal with the worsening radio-frequency-interference (RFI) environment. 
Many new surveys are also designed to maximise telescope efficiency and therefore are planned to carry out commensal spectral line and high-time resolution surveys. Such simultaneity requires the development of new observing and calibration strategies, including the use of a calibration noise source being switched throughout the observations (see, e.g., \citealt{Li+18IMMag}). 

High-time resolution surveys have historically led to serendipitous discoveries. Such discoveries include pulsars, and FRBs -- the first FRB was detected in the archival data from a pulsar survey of the Small Magellanic Cloud \citep{Lorimer+07Sci}.
There may be other signals within the existing archival data sets \citep{Zhang+20ApJS} that have not yet been identified. We know that the signatures of flare stars (and other transient signals) are likely to be present \citep{Zic+19MN, Osten08apj}, but have not yet been identified because the primary algorithms used are specific to pulsar- and FRB- type signals.  One of the biggest challenges facing the planning of the next generation of surveys is in determining how to find the ``unknown unknowns'' in massive data volumes. The Australian Square Kilometre Array Pathfinder (ASKAP) telescope has commenced widefield imaging surveys, such as the Widefield ouTlier Finder (WTF project), which is already finding unexpected source types  (e.g., \citealt{Norris+21PASA}). Similar searches for extraterrestrial intelligence (SETI) are being carried out in high-time resolution data sets \citep{Lacki+21ApJS}. Very recently, the Parkes Breakthrough Listen observations detected a suspicious signal (`blc1') at around 982\,MHz towards Proxima Centauri \citep{Smith+21NatAs}, which turned out to have terrestrial origin after detailed analysis \citep{Sheikh+21NatAs}. 

Many traditional machine learning (ML) algorithms rely on training data sets of labelled events, which are not easily obtained for unknown or rare signals. Numerous algorithms have been proposed for anomaly detection, such as edge \citep{Zhang+16arXiv} and Out-of-Distribution \citep{Yang+21arXiv} detection methods. Such algorithms can be tested on existing data sets, but real observations are complicated as they are affected by instrumental issues and RFI and the number of unexpected events within a given data set is unknown and likely to be small.  

To aid in the development and testing of new algorithms we have developed a software package, \textsc{simulateSearch}, which can simulate signals of interest (from known astronomical sources like pulsars, to RFI and more generic signals of arbitrary form) and inject those signals into actual or simulated observations. This code can be used for numerous purposes including i) developing ML training data sets, ii) determining the effectiveness of different algorithms for different source types, iii) producing data sets to allow pipelines to be developed and tested, iv) determining optimal frequency bands for selection for observatories affected by strong RFI and  v) testing the completeness of processing pipelines.

The structure of this paper is organised as follows. First, we provide an overview of the simulation software in \SEC{sec:overview}. We then show how the signals from different sources can are modelled (see \SEC{sec:simu_srcs}).  These signals can be injected into simulated data sets of a specific telescope and survey (\SEC{sec:simu_tel}) and also into archival data sets (\SEC{sec:inject}). To conclude, we discuss the use of our simulation software in \SEC{sec:disc}.


\section{Overview of the simulation software}
\label{sec:overview}

In this paper we do not provide detailed installation or usage instructions for the software. Instead, we describe the purpose of the software and detail of the algorithms implemented. The software and documentation can be obtained from \url{https://bitbucket.csiro.au/scm/psrsoft/simulatesearch.git}. The input files used to produce the first image in this paper are described in Appendix~\ref{sec:makeImages}.

The simulation software is divided into two parts. The first part produces a set of data files that record the simulated signals with high dynamic range.  The second part combines those data files accounting for telescope pointing positions (i.e., some signals will always be present, such as RFI detected in the far side lobes of the telescope, whereas astronomical sources will only be seen at a specific pointing direction) and produces data files that mimic the output data products from a telescope observing system.  We create PSRFITS \citep{Hotan+04PASA} search mode data files with 1, 2 or 8-bit quantisation and a single polarisation channel (total intensity). The observed signal is simulated as being detected (i.e., not the raw voltage data streams), channelised (into a specific number of frequency channels) and sampled with typical sampling rates ($\mu s$ to ms).

\begin{figure}
    \centering
    \includegraphics[width=0.55\textwidth]{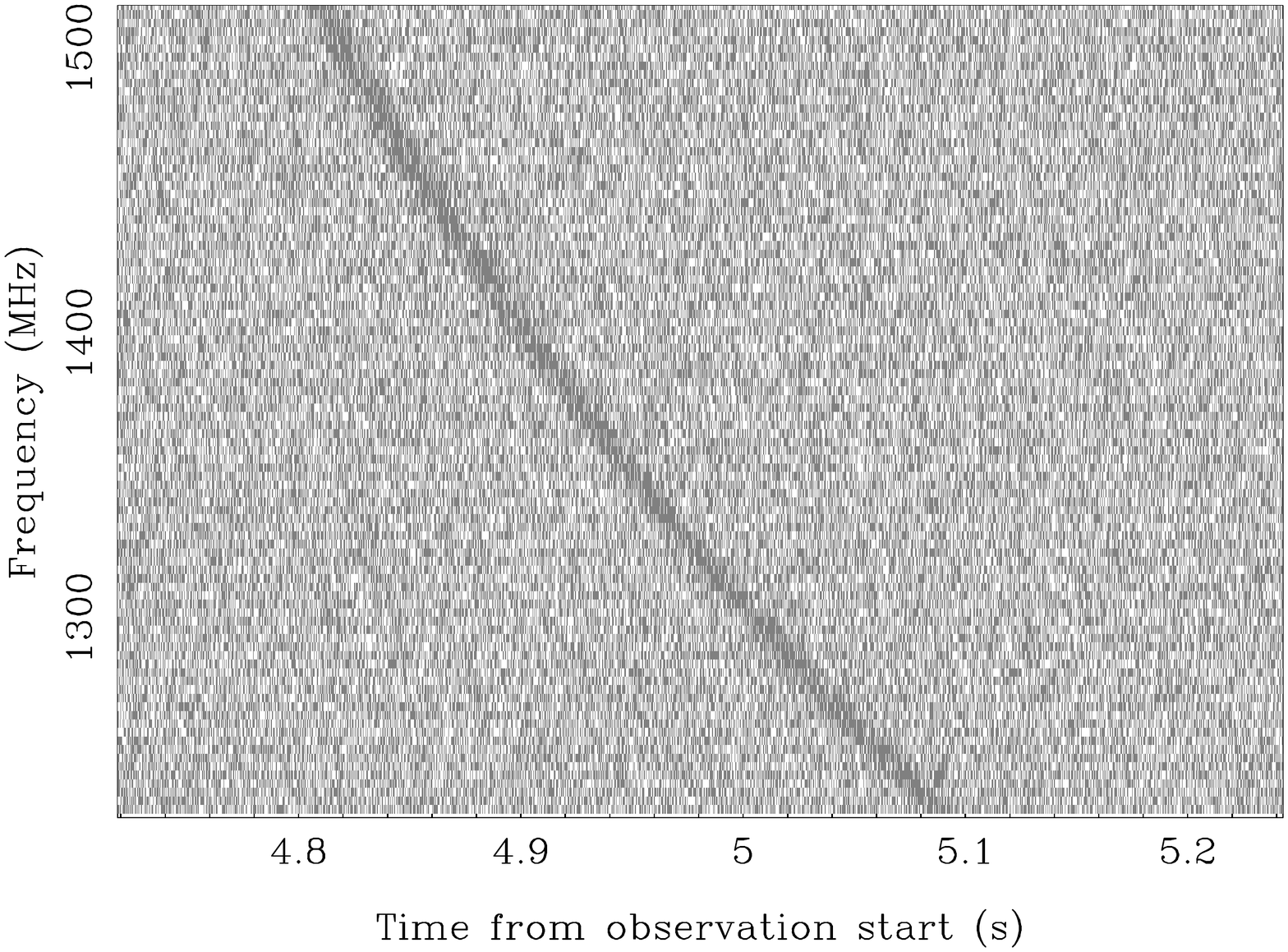}
    \caption{A simulated fast radio burst with dispersion measure of 300\,$\cmpc$. This is a mock Parkes multibeam system that is 1-bit sampled with 96 frequency channels. The system noise and the FRB event were simulated using the commands \texttt{simulateSystemNoise} and \texttt{simulateBurst} in \textsc{simulateSearch}, respectively.}
    \label{fig:frb}
\end{figure}

The output PSRFITS search mode data files can be processed by software packages for high-time resolution data processing, such as \textsc{pfits}\footnote{https://bitbucket.csiro.au/projects/PSRSOFT/repos/pfits/browse} \citep{Hobbs21ascl}, \textsc{dspsr}\footnote{http://dspsr.sourceforge.net/} \citep{vS&B11PASA} and \textsc{presto}\footnote{https://github.com/scottransom/presto}\citep{Ransom01PhDT}. An example is shown in \FIG{fig:frb} where we have simulated a ``dispersed pulse'' in a dynamic spectrum, exhibiting the intensity of radio signal as a function of frequency and time.
This specific simulated data file has been quantised using a single bit and thus each time and frequency sample is either 0 or 1. Such files can be processed as if they were from actual observations. Typically the data processing starts by de-dispersing the data sets at a range of trial dispersion measures and then searching the de-dispersed time series for impulsive or periodic signatures.

Splitting the simulation process into two parts allows the final data products to be produced for multiple telescope systems or observing strategies. For instance, the same radiometer noise may be included in all output data products with a variable-amplitude simulated astronomical source. In another use-case, the same astronomical signal may be present, but the user may wish to trial their algorithm with different levels of radiometer noise.  

The format of the data files containing the simulated signals is relatively simple and described in \APP{app:format}. Our software contains various routines for simulating commonly-used signals, but the user can also produce new simulated signals relatively easily in Python, C or other languages. The simplest form of the simulated signal data files is a binary representation of the signal for the entire data span recorded as  32-bit floating point values. Long duration simulations with a large number of frequency channels can lead to extremely large data files. For a single FRB event, which may only last for a millisecond in an observation lasting hours, storing the source signal for the entire survey length is clearly unnecessary. The software therefore provides methods to compress the source signals (details are provided in \APP{app:format}).


\section{Simulating the sources}
\label{sec:simu_srcs}
\subsection{Radiometer noise}

Radio telescope observations are affected by radiometer noise.  We model the frequency-dependent system noise as being drawn from a Gaussian distribution with amplitude defined by the radiometer equation: 
\begin{equation}
	S_{\nu, \mathrm{rms}}=\frac{T_{\mathrm{sys}}}{G_{\rm dpfu}\sqrt{N_\mathrm{p}\,\Delta\nu\,\tau}}
	\label{eq:rmeq}
\end{equation}
where $G_{\rm dpfu}$ is the telescope gain in degrees per flux unit ($\rm K/Jy$), $T_\mathrm{sys}$ is the system temperature, $\tau$ the digital sampling time and $\Delta\nu$  the receiver bandwidth. Throughout the simulation software we assume that the signals and noise are represented in units of janskys.

\begin{figure}
    \centering
    \includegraphics[width=0.5\textwidth]{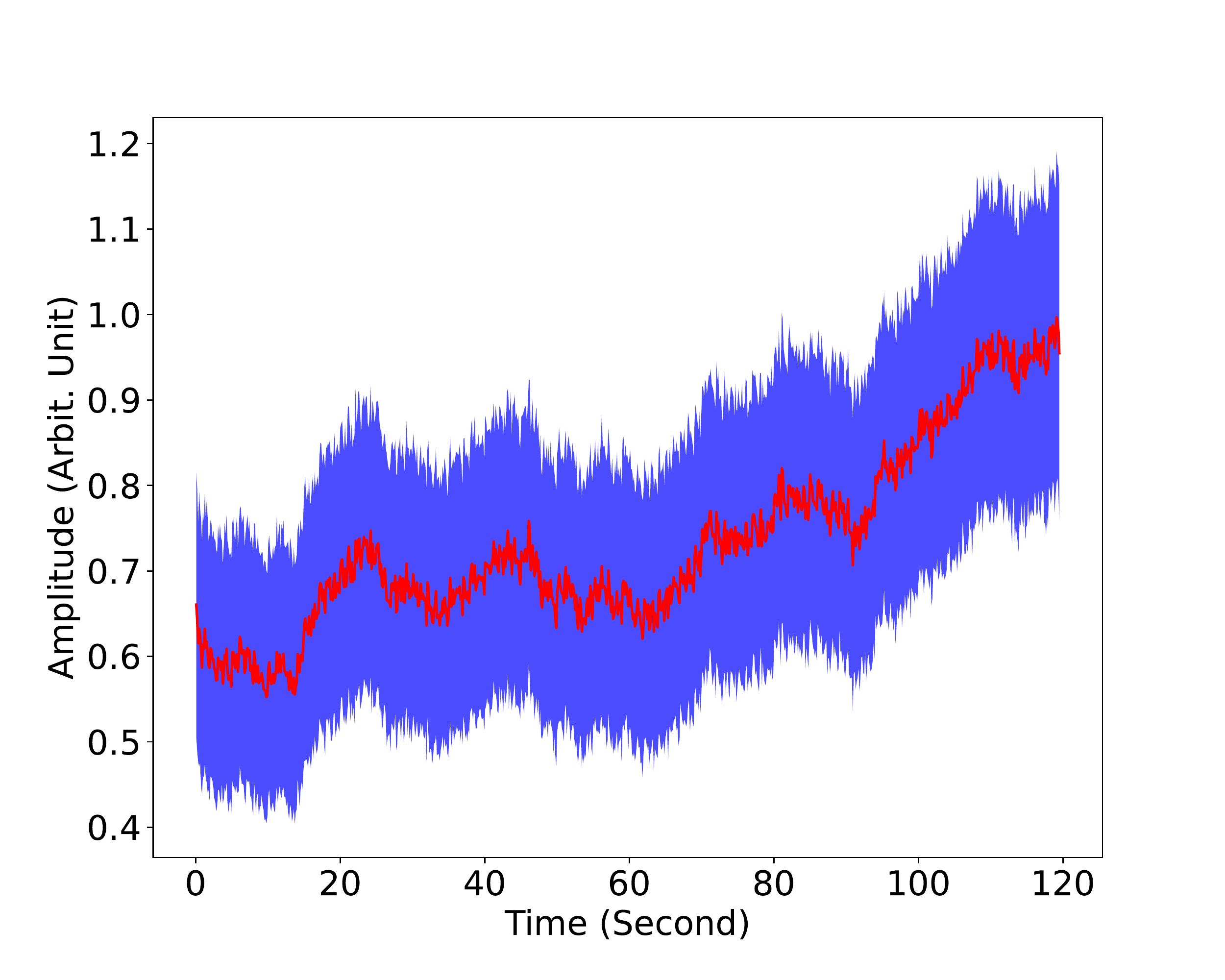} \\
    \includegraphics[width=0.5\textwidth]{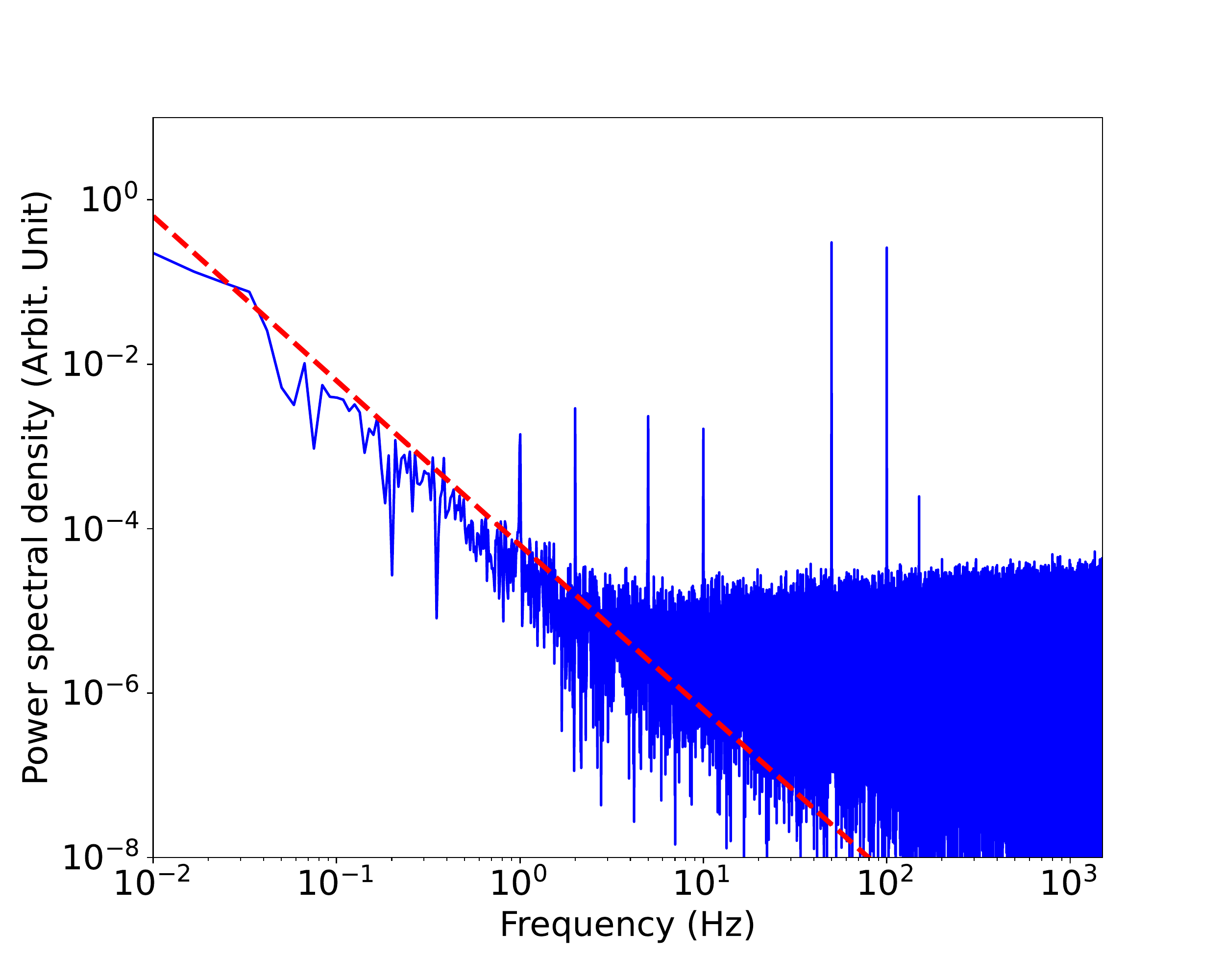} 
    \caption{The non-dispersed time series of the noise data (upper panel), and the power spectrum of simulated radiometer noise and low-frequency red noise (lower panel). The data has been simulated using \texttt{simulateSystemNoise}. In the upper panel, the mean of simulated noise is denoted as red curve and its standard deviation is shaded in blue. In the lower panel, the spectral slope of red noise background is $\alpha=-2$ (red dashed line), six sinusoidal tones have been injected at 1, 2, 5, 10, 50, 100 and 150\,Hz using \texttt{simulateRFI} to represent ``birdies''. }
    \label{fig:noise}
\end{figure}

High-time resolution data sets can be affected by low-frequency noise processes.  Such noise can arise from gain variations, telescope-pointing jitter, spill-over variations, etc. Such noise can impact on the detection of radio transients like FRBs (e.g., \citealt{Zhang+21MN}) and requires de-reddening procedures in pulsar searches (e.g., \citealt{Lazarus+15ApJ, Suresh+22arXiv}). We provide the ability to model red noise with power spectral density
\begin{equation}
    S_\mathrm{red} = S_0 f^{\alpha}.
\end{equation}
Here $\alpha$ is the spectral index and $S_0$ the power spectral density at 1\,Hz.  Our code also allows for a cut-off frequency in the simulated red noise (the modelled power spectral density below the cut-off is zero) or a frequency at which the spectrum flattens (the power spectral density below the flattening frequency equals the power at that frequency). An example is shown in \FIG{fig:noise} where red and white noise has been simulated, the observing-frequency channels summed to form a time series and then Fourier transformed.

\begin{figure*}
    \centering
    {\includegraphics[width=20cm]{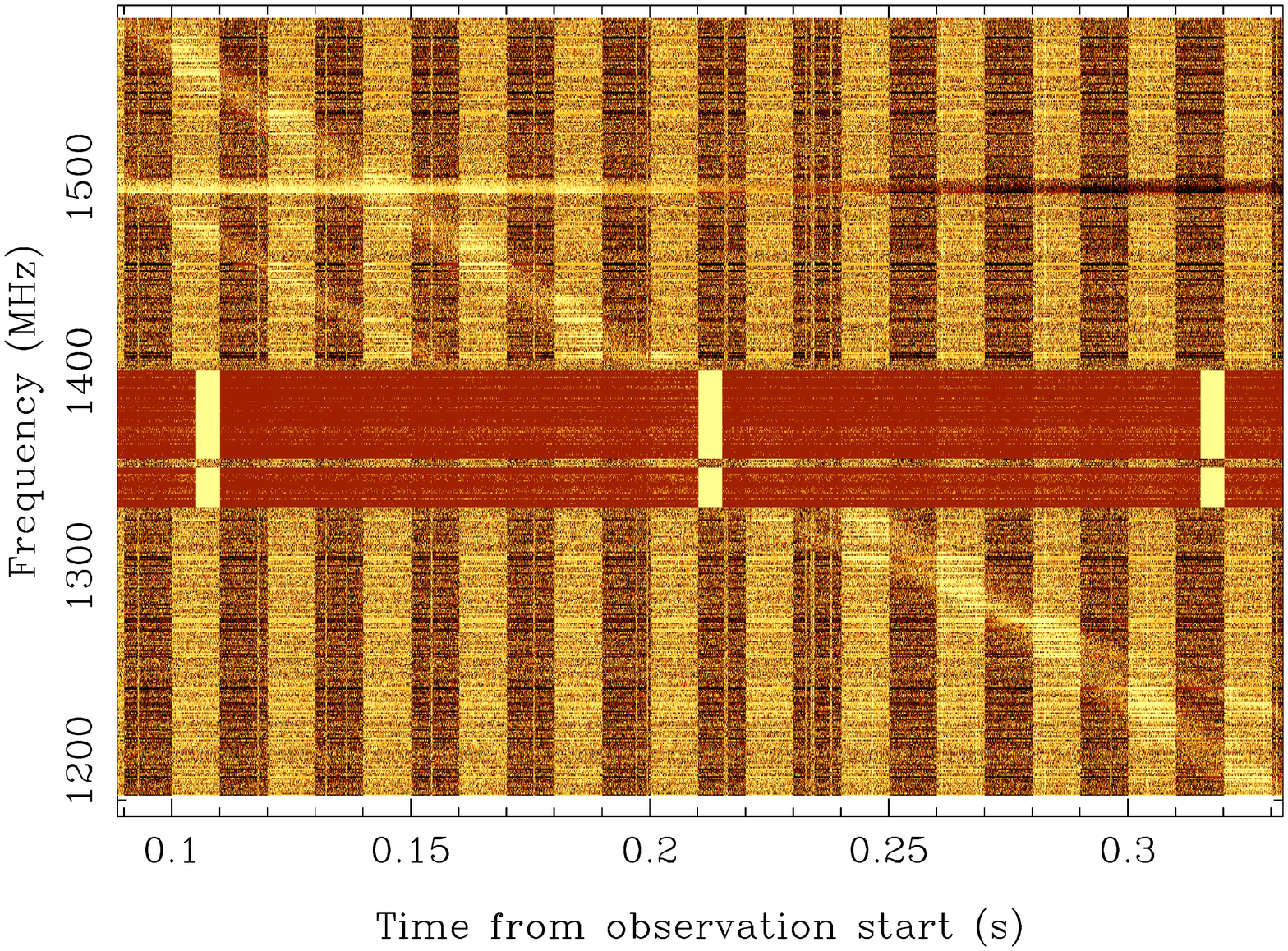}}
    \caption{Simulated data set that includes the signature of a flare star (relatively weak broadband drifting signal), impulsive RFI (vertical, narrow stripes), satellite and terrestrial tower RFI (horizontal stripes with varied bandwidths) and a switching noise source (the dark/light vertical structure). The data set has been sampled every 64$\mu$s with 512 frequency channels and 2-bit samples.}
    \label{fig:rfi_mix}
\end{figure*}

\subsection{Radio frequency interference}

Radio frequency interference (RFI) is defined here as unwanted radio signals in the astronomical data sets (which may be self-generated at the observatory or from external sources). RFI is ubiquitous and dealing with RFI is a major challenge in current radio astronomy. We can divide RFI into three categories: 1) impulsive and wideband RFI, 2) narrowband and persistent RFI and 3) birdies (which produce periodic signals across the band from, for instance, the mains electricity supply).  All these types of RFI can be simulated using our software.  

Impulsive, broadband RFI is simulated based on a distribution of amplitudes and pulse widths detected at a specific observatory.  The events are randomly distributed in time through the simulation.  We used an analysis of the zero dispersion measure, impulsive RFI detected in representative observations of the High Time Resolution Survey (HTRU; \citealt{Keith+10MN}) to determine the typical amplitudes and widths of impulsive RFI detected at the Parkes observatory. Such events are shown as the narrow (only one or two time-samples wide), vertical stripes in \FIG{fig:rfi_mix}.


We can provide a simple simulation of the signals from transmission towers by increasing the system noise significantly in frequency bands where there is a strong transmitter. Examples are shown as the horizontal, relatively narrow-band features in  \FIG{fig:rfi_mix}. The detectability of any astronomical source (such as the flare star simulated in this data set) in such bands will be significantly reduced (and often undetectable).  

Most RFI is time dependent; either through intrinsic time variability or because the telescope pointing direction is not constant with respect to the location of the interfering source.  We do not attempt a detailed simulation of all known types of time variable RFI. For instance, mobile transmission towers and handsets, WiFi, BlueTooth, aircraft navigational devices all produce strong signals that are time variable.  Our simulation code allows for relatively simple representations of such signals.  One example is shown near the centre of \FIG{fig:rfi_mix} and represents a simplified representation of the signals observed from a mobile transmission tower.

\begin{figure}
    \centering
    {\includegraphics[width=10cm]{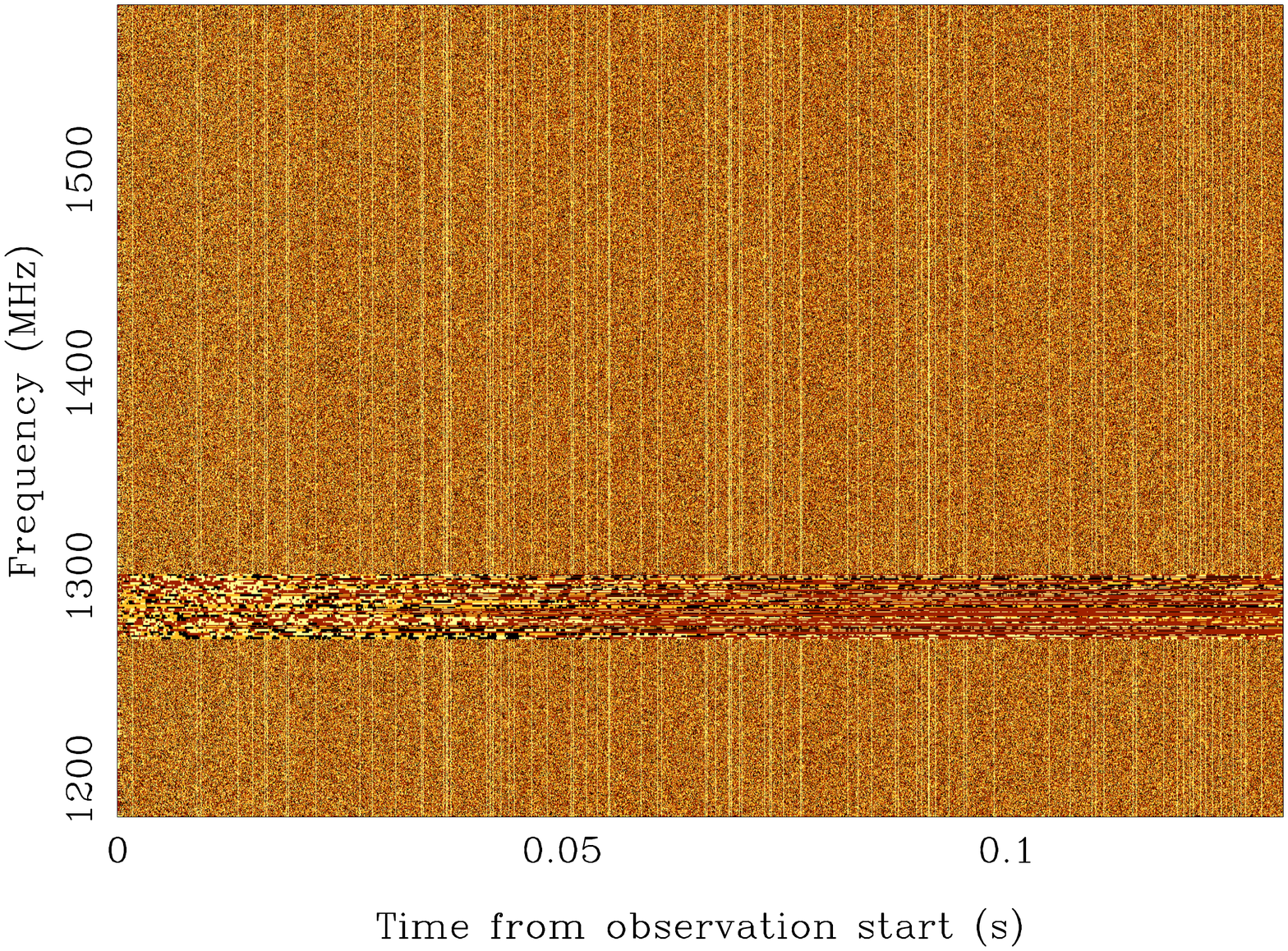}}
    \caption{Simulation of the Long Term Evolution Network (4G LTE) and how impulsive RFI specific to one part of the observing band (around 1290 MHz) can affect the entire digitised data set (producing the vertical stripes). This image was made using \texttt{simulateLTE} and \texttt{createSearchFile}.}
    \label{fig:lte}
\end{figure}

The Parkes ultra-wide-bandwidth receiver system \citep{Hobbs+20PASA} is affected by RFI. Some of the most problematic are mobile transmission towers that produce impulsive RFI that can be extremely bright for short durations.  Analogue-to-digital converters are close to being limited  by the jitter noise of the sample clock (see Tuthill et al., in prep.).   This leads to power in one part of the digitised band being ``smeared'' across the entire band.  The Parkes data sets are affected by this issue and we are trialling various methods to mitigate this effect. In \FIG{fig:lte} we show a simple simulation of 4G mobile transmission, which is highly impulsive in small sub-bands.  We smear a tiny fraction ($10^{-8}$) of the signal strength across the entire band, which produces a signal that is similar to our actual observations in some sky directions.

Signals from satellites are a major RFI source at all observatories. The positions of individual satellites can be predicted using two line element (TLE) sets. However, for many of the satellites (such as the global positioning satellite systems), there are a large number of satellites above the horizon at a given time and the signal from each satellite is detected in the far side lobes of the telescope beam. To simulate such sources we assume that each satellite emits multiple signals each of which follows a sinc$^2$-frequency-dependent function:
\begin{equation}
    P(f)=A \left[\frac{\sin[a(f-f_0)]}{a(f-f_0)}\right]^2
\end{equation}
where $f$ is the observing frequency and $f_0$ is one of the emission frequencies from the satellite.  Instead of attempting to model the structure of the far side-lobes of any particular telescope, we assume that the motion of the source through the far side-lobes of the telescope will introduce a sinusoidal variation in the signal amplitude:
\begin{equation}
    A = A^\prime + A^\prime\sin(2\pi t /P + \phi)
\end{equation}
where $A^\prime$ is the signal amplitude, $P$ is the variability timescale and $\phi$ is the signal phase. We show an example (close to 1500\,MHz) in \FIG{fig:rfi_mix} of the RFI caused by the global positioning satellites. 

To simulate tones, we inject a sinusoidal signal (or, if requested, a rectified sinusoidal signal) with a given amplitude, phase and frequency. Such signals are typically only apparent after a periodicity search has been carried out on the final data product and examples of common signals are shown in \FIG{fig:rfi_mix}.  In contrast some recent (or planned) surveys include strong periodic signals that are purposely injected into the data stream. This is because commensal high-time resolution and high frequency resolution surveys are now being carried out in order to maximise telescope efficiency by searching for pulsars, FRBs and spectral lines simultaneously (for instance, the CRAFTS survey being carried out with FAST; \citealt{Li+18IMMag}).  Spectral line surveys  traditionally make use of a switched noise source throughout their observations to track and account for telescope gain variations and for amplitude calibration. The simulation code can therefore include such a broad-band, switching noise source and we provide an example in \FIG{fig:rfi_mix}.

\subsection{Transient astronomical signals}

\begin{figure}
    \centering
    \includegraphics[width=0.55\textwidth]{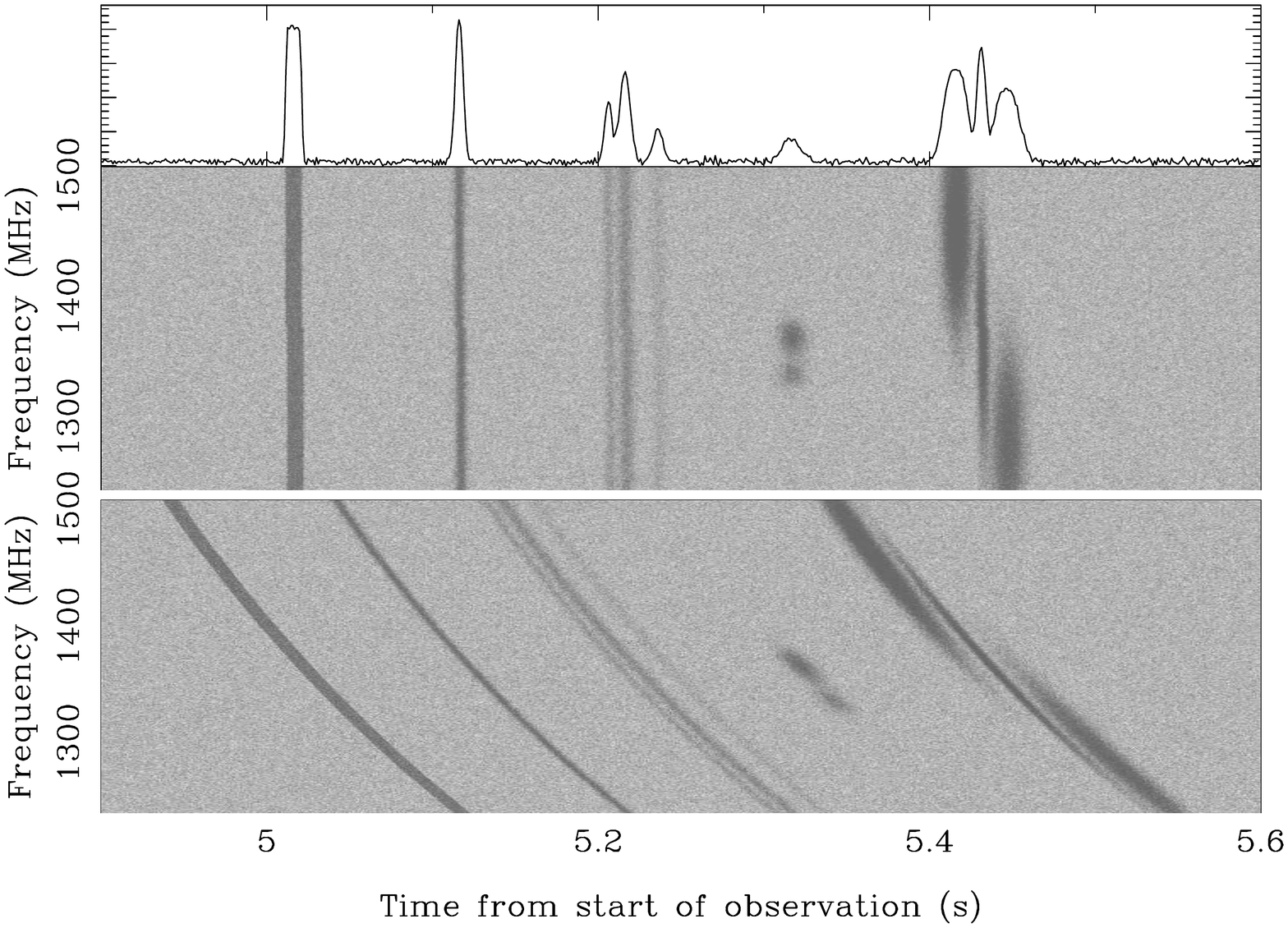}
    \caption{Examples of individual burst events.  The bottom panel shows the simulated data set. The bursts all have a DM of 200\,$\cmpc$.  The central panel shows the de-dispersed data sets and the top panel the frequency-summed time series. The events were simulated using \texttt{simulateBurst} and represent from left to right: (1) an event with rectangular edges, (2) a burst with a Gaussian profile, (3) a multi-component event, (4) an event with band-limited spectrum and (5) a burst with downward frequency drift.}
    \label{fig:profile}
\end{figure}

We know that high-time resolution data sets will include individual bright pulses from pulsars, fast radio bursts (FRBs) and related sources such as rotating radio transients (RRATs). Such transient events are parameterised using an event time, pulse amplitude, pulse width and dispersion measure (DM). The pulse shape can be simple (modelled using a single Gaussian or rectangular profile) or defined using multiple Gaussian components. We can also build band-limited \citep{Kumar+21MN} and downward-drifting structure \citep{Hessels+19ApJ} in the burst spectrum. Examples are shown in \FIG{fig:profile} for bursts with different profile and frequency structure.

The aim of many high-time resolution surveys is to find fast transients (such as pulsars and FRBs).  The use of low-bit quantisation in the output data products has required level setting procedures to account for longer term system noise variations (more details are provided in Section~\ref{sec:process}).  This has ruled out the chance of finding slower transient signals.  However, several classes of stars across the Hertzprung-Russell Diagram, including the Sun, produce intense bursts of non-thermal radio emission, powered by various forms of magnetic activity \citep[e.g.,][]{Guedel02ARA&A}. These include auroral activity \citep{Zarka98jgr,Hallinan15nat,Trigilio11apjl}, coronal activity, driven by flaring, space weather, and other dynamic processes within stellar atmospheres and astropheres \citep[e.g.,][]{Bastian90solphys,Pick08aapr,Benz17LRSP}; and the interaction of magnetic fields between two components in close binary systems such as RS Canum Venaticorum systems \citep{Drake89apjs,Toet21aap}. Regardless of the driving mechanism, active stars can produce radio emission variable on timescales from milliseconds \citep[e.g.,][]{Osten08apj} to days \citep{Slee03pasa}, with complex time-frequency structure. The ability to capture this variability across such a broad range of timescales remains under-explored. As such signals are likely to exist in our current archival data sets, yet the pulsar-based algorithms developed so far are unlikely to detect them, we have included the ability to simulate likely flare-star signatures.  We parameterise such events as relatively broad-band sources covering a specified bandwidth, where individual components drift in time and frequency following a quadratic polynomial (allowing both linear and quadratic drifts).  An example is shown in \FIG{fig:rfi_mix}.

\subsection{Periodic pulses}

Pulsars produce a sequence of periodic pulses. Apart from the brightest known pulsars, the majority of these pulses are so weak that they cannot individually be detected.  Instead the search-mode data streams are de-dispersed, averaged across frequency channels and then Fourier transformed to search for periodic signatures within the data.  

\begin{figure}
    \centering
    \includegraphics[width=0.6\textwidth]{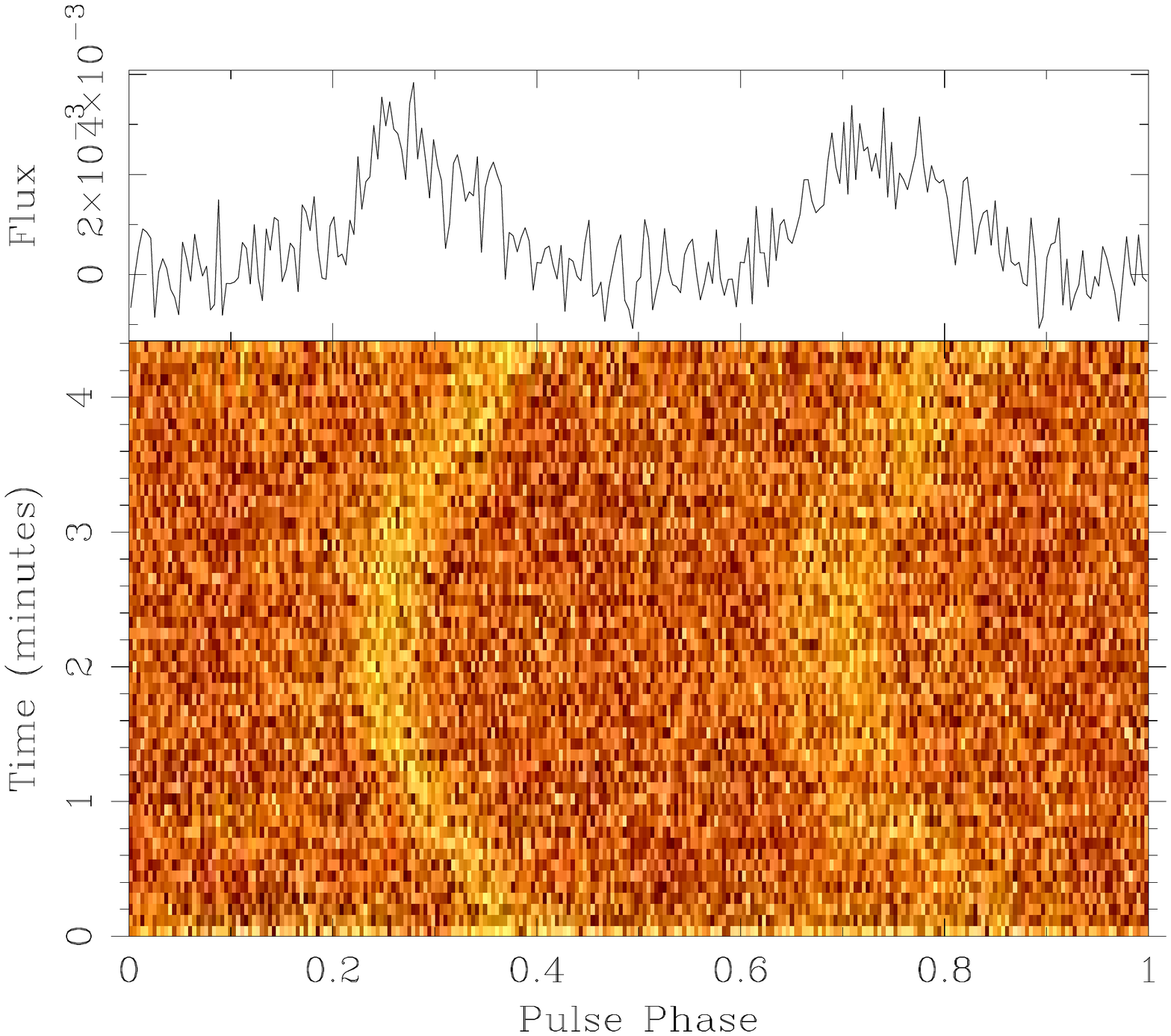}
    \includegraphics[width=0.6\textwidth]{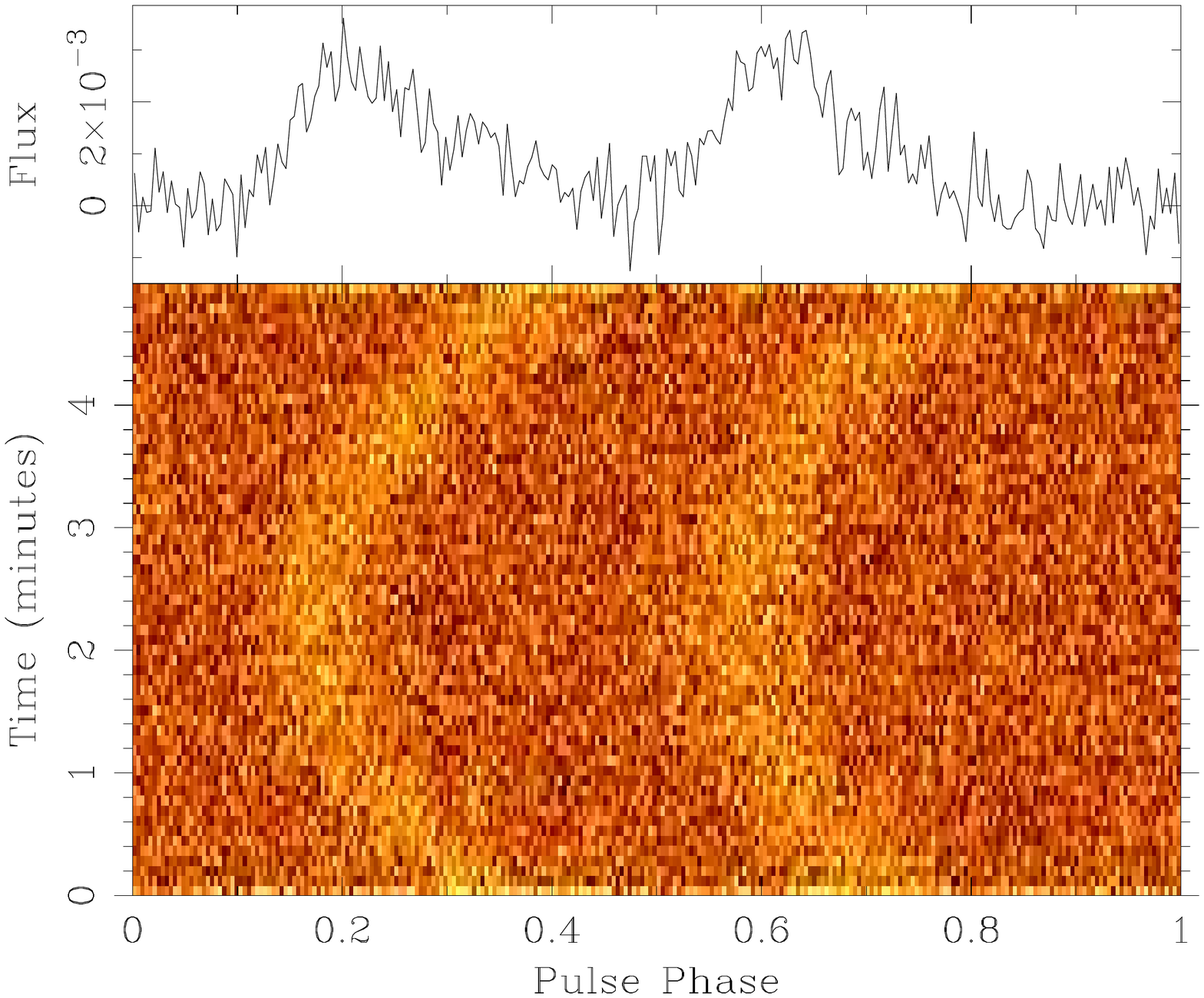}
    \caption{The real (top panel) and simulated (bottom panel) folded data for the highly relativistic pulsar, PSR J0737-3039A. In each panel the sub-figures contain the folded pulse profile over the entire observation duration (top sub-panel) and the bottom sub-panel shows the profile as a function of time throughout the observation. The real data file \textsc{PH0042\_004B1.sf} was obtained from the CSIRO DAP via \url{https://doi.org/10.4225/08/598c2d9103f0c}. The data were folded and plotted using \textsc{dspsr} and \textsc{psrchive} routines. The simulation was carried out using \texttt{simulateComplexPulsar} and \texttt{createSearchFile}.}
    \label{fig:double_psr}
\end{figure}

We simulate pulsars using a \textsc{tempo2}-style predictor \citep{Hobbs+06MN}  that can be used to determine the arrival times of pulses at a specific observing frequency and specific observatory.  The use of predictors allows highly relativistic binary systems to be modelled.  For instance, in \FIG{fig:double_psr} we compare an actual observation with the simulation for PSR~J0737$-$3039A, which was discovered by Parkes high-latitude multibeam pulsar survey \citep{Burgay+03Nat} and is a highly relativistic binary system \citep{Kramer+21PhRvX}. The real data set was obtained from the CSIRO Data Access Portal\footnote{https://data.csiro.au/collections} (DAP, \citealt{Hobbs+11PASA}) and we folded the data at a nominal rotational period (upper panel in \FIG{fig:double_psr}).  Our simulated result (lower panel) is based on a \textsc{tempo2} predictor from an up-to-date timing ephemeris for this pulsar. 

The pulse profile can be modelled using multiple Gaussian components. The resulting pulse shape is subsequently rescaled to ensure the area represents the mean flux density of the pulsar. Each individual pulse from the pulsar is simulated.  We therefore can vary the intensity of each individual pulse.  As the intensities of individual pulses are drawn from various distributions (as described for terrestrial use by \citealt{Dawson+22A&C}), we allow the user to provide their own distribution that is then used within the simulation.



\subsection{The interstellar medium}

The observed astronomical signals have propagated through the interstellar medium (ISM).  The time delay caused by dispersion for a pulse measured at two different frequencies of radio waves ( $\nu_1$ and $\nu_2$) is given by:
\begin{equation}
\Delta t =4.15\, {\rm ms}\, \left(\frac{\DM}{1\, \cmpc}\right)
	\left[\left(\frac{\nu_1}{1\, \mathrm{GHz}}\right)^{\alpha} -\left(\frac{\nu_2}{1\, 
	\mathrm{GHz}}\right)^{\alpha}\right].
	\label{eq:timfreq}
\end{equation} 
The dispersion measure (DM) is defined as the integral of electron density along the light of sight,
\begin{equation}
	\DM=\int n_{\rm e}\, \D l \,.
	\label{eq:dm}
\end{equation}
Typically $\alpha = -2$ for cold, diffuse, ionised gas.  Dispersion is simulated in our software, where $\alpha$ and DM can be defined by the user. Note that negative dispersion measures are permissible, as they may be produced by astronomical signals with intrinsically negative drift. 

The signals are also affected by scintillation. Various models for scintillation are possible and our simulation software allows the user to develop their own models of scintillation as required. Here we consider only diffractive scintillation in the case of strong scattering. For given scintillation time-scale and bandwidth, we simulate a dynamic spectrum following procedures described in \cite{Dai+16MN}. In the dynamic spectrum the flux density as a function of time and frequency is $S_{\rm dyn}(t, \nu)$.   We determine this from a 2-dimensional auto-covariance function defined using a user-provided scintillation timescale ($\tau$) and bandwidth ($\Delta \nu$) \citep{Dai+16MN}. 

We note that FRB signals often have frequency-dependent amplitude fluctuations that cannot be modelled through scintillation.  We therefore provide options for the user to specify any frequency evolution of the pulse events.  An example of a FRB that is constrained to a small frequency range is shown as the event of example (4) in \FIG{fig:profile}.

\begin{figure}
    \centering
  \includegraphics[width=0.55\textwidth]{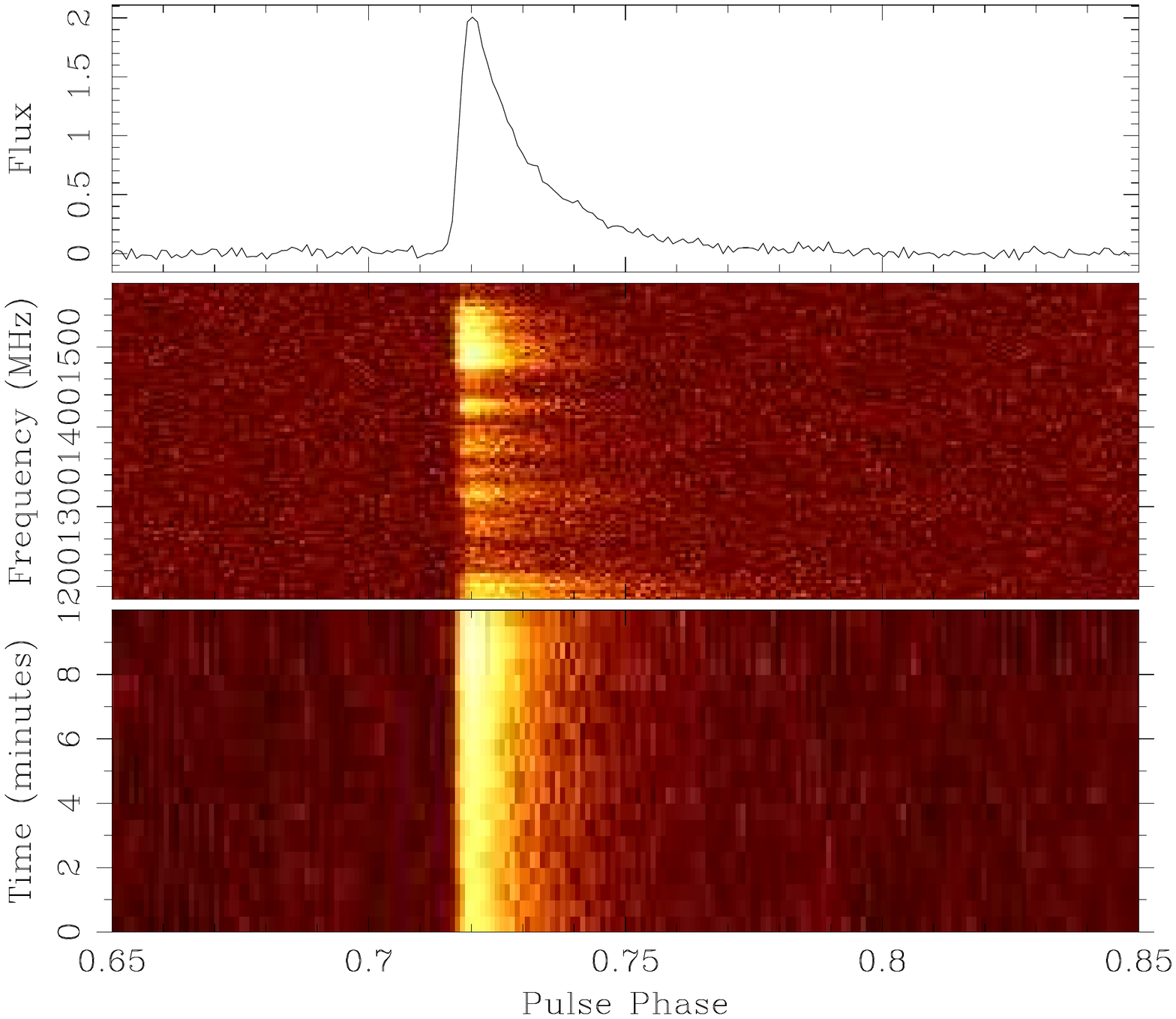}
    \caption{Simulation on the propagation effects of pulsar observation. A dynamic spectrum was simulated. We also added in the effect of scattering.  The top panel shows the time- and frequency-averaged pulse profile. The central panel shows the frequency structure after averaging in time and the bottom panel shows the time structure after averaging in frequency.  Note this is to provide an example of these effects and we have not tried to link the scattering and scintillation properties of a real pulsar.}
    \label{fig:ism}
\end{figure}

Pulse events also undergo scattering in the interstellar medium. We therefore provide the ability to convolve a pulse signal with an exponential function with a time scale defined by a specified DM. We use the pulse-broadening function in \cite{Bhat+03ApJ} to model the scattering effect on integrated pulse profile,
\begin{equation}
    g(t)=\frac{\exp(-t/\tau_\mathrm{d})U(t)}{\tau_\mathrm{d}},
\end{equation}
where $\tau_\mathrm{d}$ is the pulse broadening time and $U(t)$ is the unit step function. To demonstrate the scintillation and scattering simulations we model a pulsar with a dispersion measure of 600\,$\cmpc$, a diffractive bandwidth of 50\,MHz and diffractive timescale of 1\,minute (we note that these are not necessarily independent parameters for actual pulsars, but choose these parameters here to allow us to demonstrate multiple ISM-related effects in a single figure). The resulting profile is shown in \FIG{fig:ism}.

\subsection{The unknown}

One aim of our simulation software is to provide tools to simulate unexpected, or unknown, sources.  We simulate such ``unknown unknown'' signals in the following three ways: (1) generalising the burst event simulations, (2) using spline curves and (3) generic images \footnote{In 1974, the active SETI used the Arecibo radio telescope to transmit a message towards the globular cluster M13. This message consisted of multiple generic images digitised in 1679 bits \citep{Atri+11SpPol}.}. 


We generalise the burst events by allowing positive or negative dispersion measures and allowing any choice of $\alpha$ in \EQ{eq:timfreq}. We also allow any user-defined frequency evolution of the burst intensity.

The high-time resolution data streams simulated here typically have relatively low frequency resolution. Therefore any unexpected source signals that will be detected in such data sets are likely to be relatively short in duration, but cover a wide band.  We therefore allow for arbitrary broadband signatures to be simulated using cubic splines, which are defined by specific time and frequency control points.  

We can embed information into the data set by producing an image (which could be obtained from the large number of available online datasets used for training ML algorithms) and using the pixels in that image  to represent the time-frequency information in the simulated data set.

\begin{figure}
    \centering
    \includegraphics[width=0.55\textwidth]{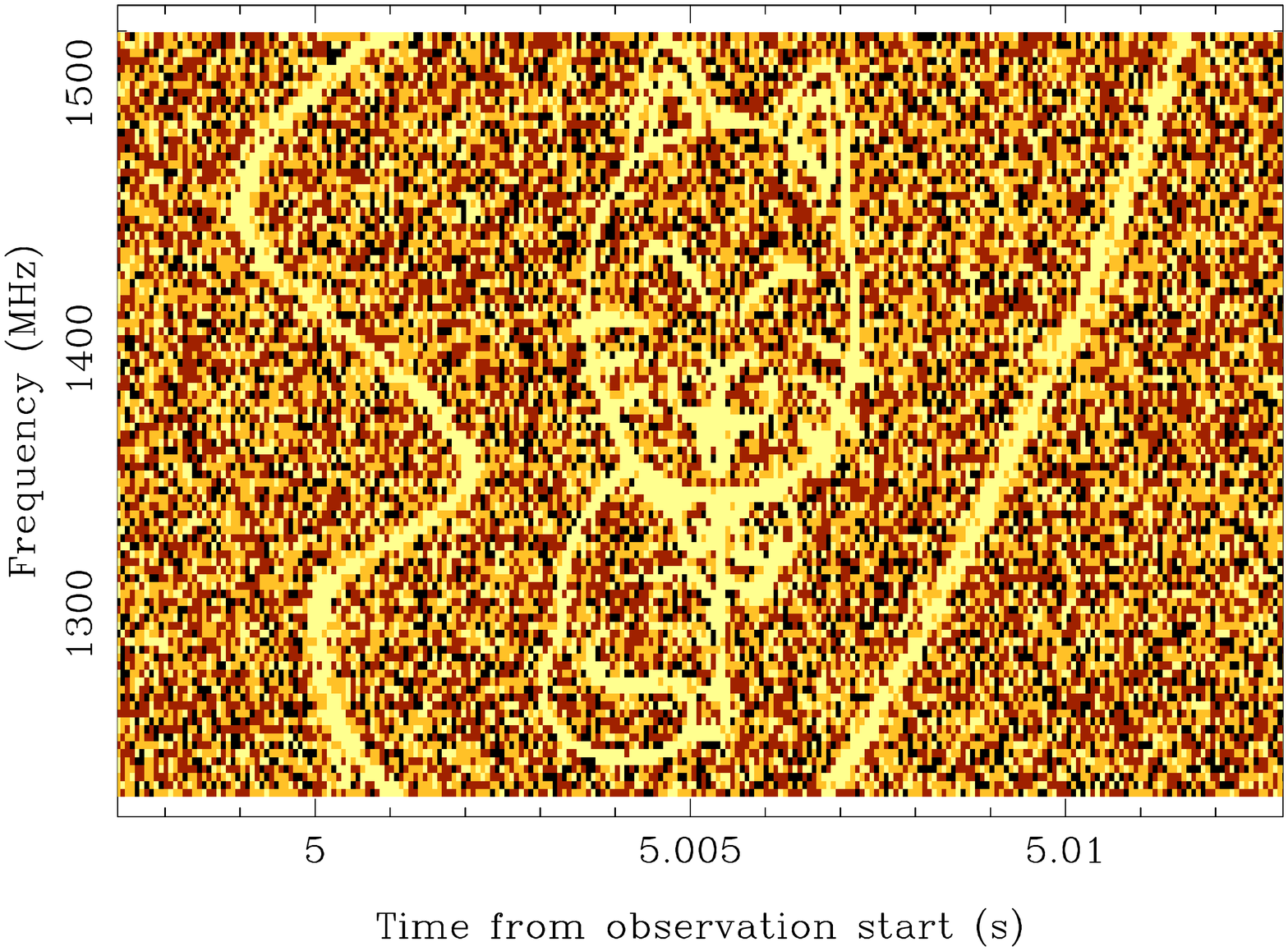}
    \caption{A simulation of unexpected signals. We use \texttt{simulateGeneric} and \texttt{simulateBurst} to inject into a 2-bit data stream an arbitrary curve defined using a cubic spline (left), an image of a cat (centre) and an FRB-like signal with a negative dispersion measure (right).}
    \label{fig:signals}
\end{figure}

We provide an example of these three types of ``unknown unknown'' signals in \FIG{fig:signals}. This contains, from left to right, an arbitrary cubic spline event, a cartoon image of a cat and an FRB-like event with a negative dispersion measure.



\section{Simulating the telescope observing system}
\label{sec:simu_tel}

In order to simulate the output data product from a high-time resolution radio survey we need to model the telescope pointing direction, the receiver system and, for each beam of the receiver, the signal path from the receiver to the astronomy data processor (a description of the signal path for a modern observing system is given by \citealt{Hobbs+20PASA}).

\subsection{The receiver system}

\begin{figure*}
    \includegraphics[width=0.48\textwidth]{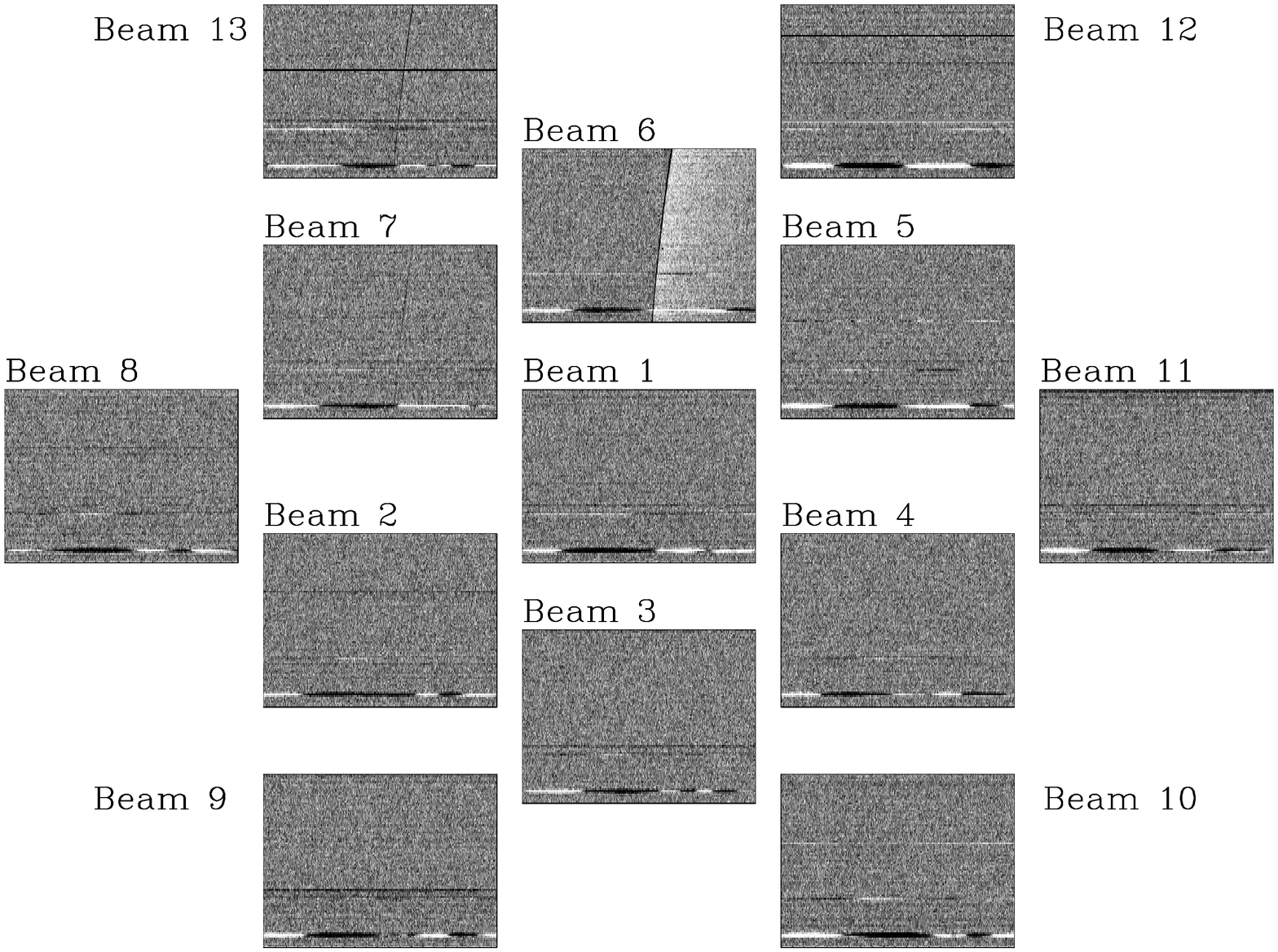}
    \includegraphics[width=0.48\textwidth]{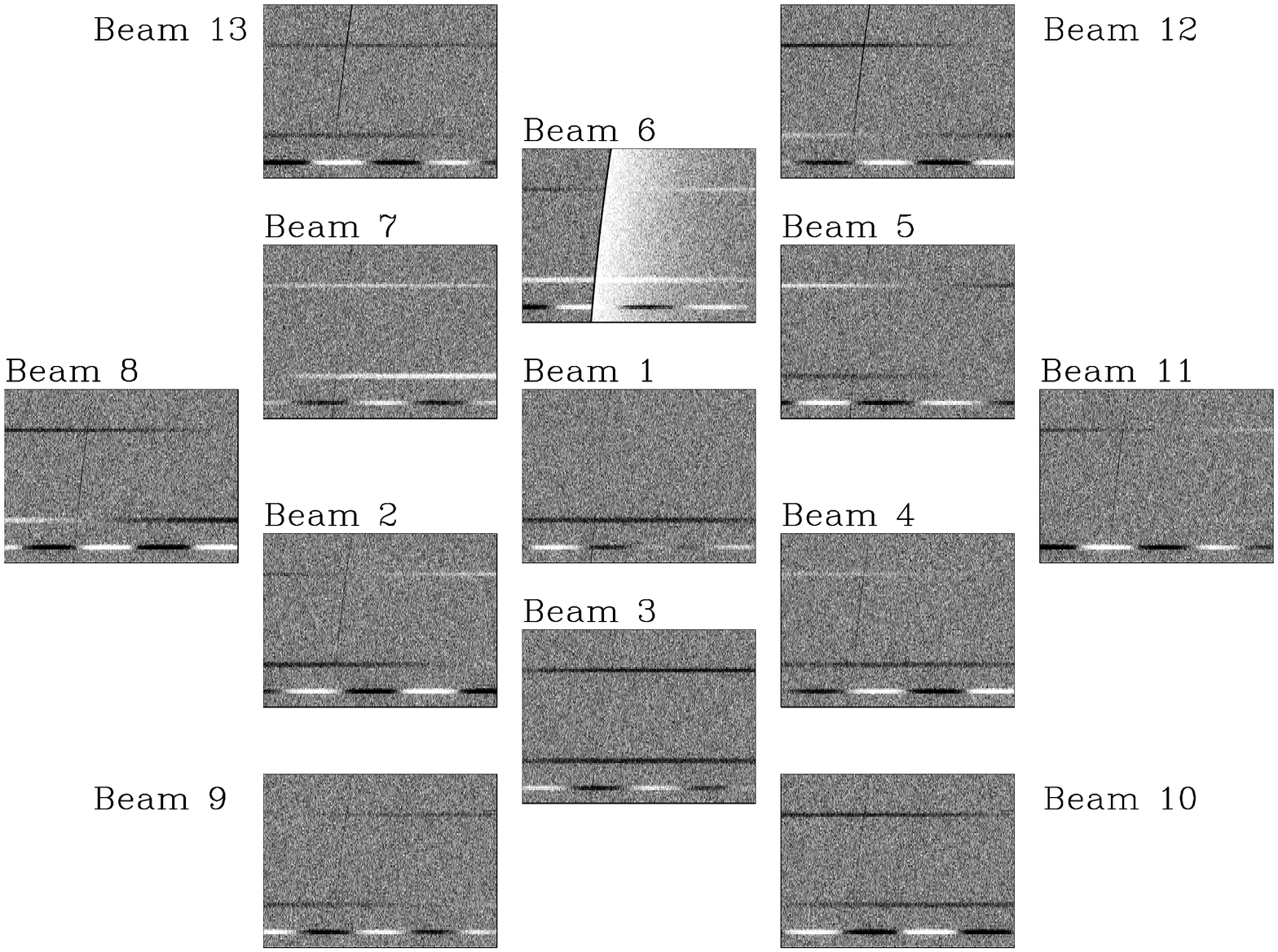}
    \caption{Multibeam view of FRB~010724 (the ``Lorimer Burst'') from the real (left) and simulated data (right). The real data files \textsc{SMC021\_008*1.sf} were obtained from \url{https://doi.org/10.4225/08/5819628e4fed9}. Each sub-panel represents time (x-axis) and frequency (y-axis) for each of the 13 beams that were simultaneously recorded, note that the frequencies range from 1231.5\,MHz (top) to 1516.5\,MHz (bottom) in the plot of each beam. The FRB is mostly clearly detected in Beam 6, but can also be seen in other adjacent beams.  The horizontal lines are the signatures of RFI from point-to-point microwave links.}
    \label{fig:lorimer_burst}
\end{figure*}

The simulated source signals are either only present in specific sky directions (different pulsars are at different sky coordinates) or are always present (for instance, radiometer noise). In a multi-beam system, each receiver beam can be modelled independently and the sky position of that beam can be defined as a function of time (allowing for simulations of scanning surveys or tracking observations of a specific sky direction). The beam pattern on the sky is given by the telescope diameter and observing frequency and assumed to follow:
\begin{equation}
    s = \left[\frac{sin(\theta)}{\theta}\right]^2
\end{equation}
where $\theta = 1.22 \lambda/D$, $\lambda$ is the observing wavelength and $D$ the telescope diameter. As an example, we show in \FIG{fig:lorimer_burst}, a simulation of the detection of the first FRB (the ``Lorimer burst'').  The left panel shows the actual data from the Parkes telescope \citep{Lorimer+07Sci} and the right panel shows our simulation.  The event is seen primarily in Beam 6, but is also present in other beams (more detectable at lower frequencies where the beam is wider). Note that we model narrow-band interference as being detectable in all of the beams.



\begin{figure}
    \centering
    \includegraphics[width=0.5\textwidth]{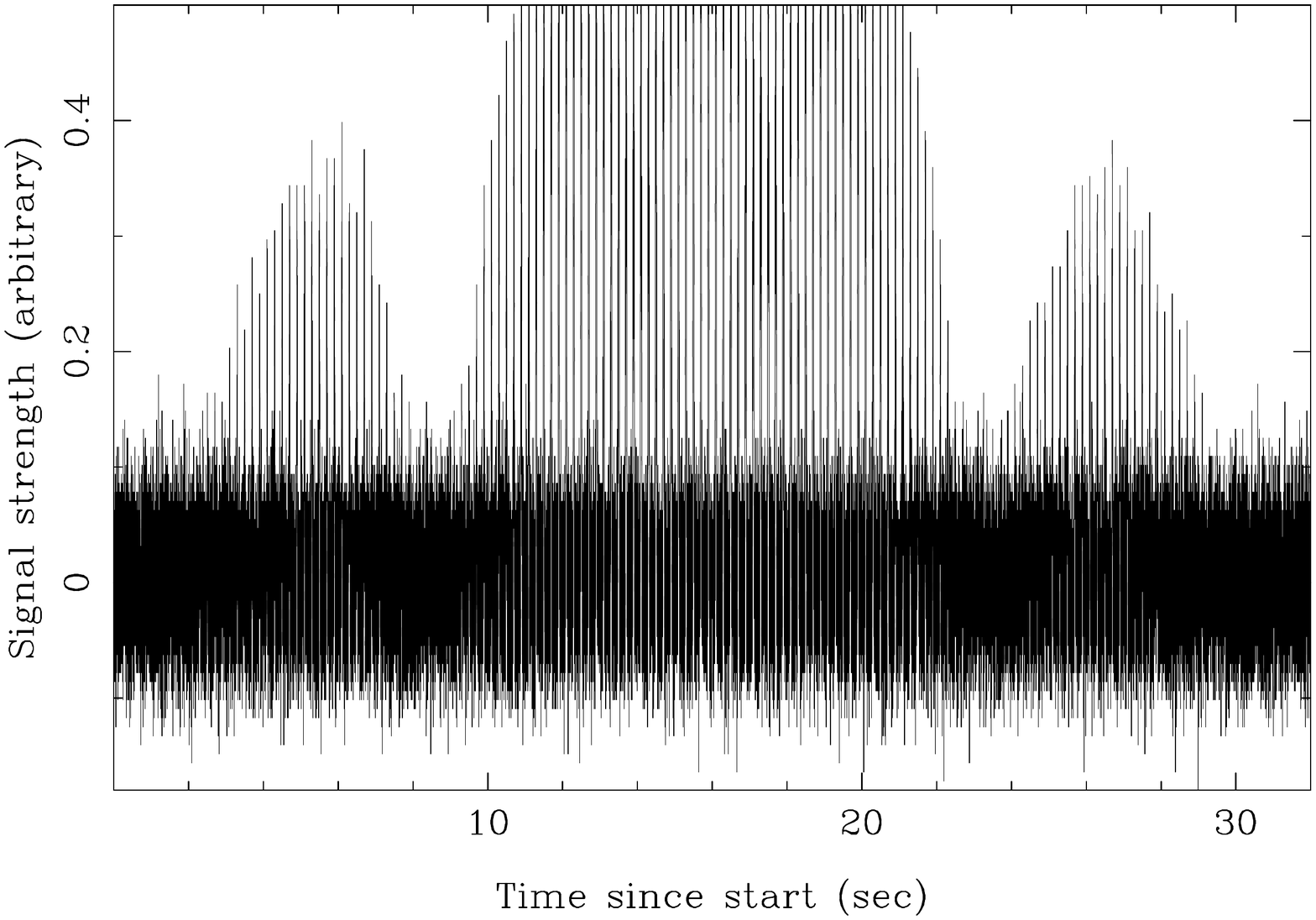}
    \caption{A representative pulse train after de-dispersion. The pulsar is simulated with a period of 0.2\,s.  We assume a large-diameter telescope like FAST observed this pulsar in a drift-scan mode. The data for this figure was simulated using \texttt{simulateComplexPsr}.}
    \label{fig:drift}
\end{figure}

Our simulation software can also model a drift-scan survey in which the telescope is held fixed and astronomical sources will drift through a beam. In \FIG{fig:drift} we show how a pulsar will be detected in such a survey.  We have assumed here a sensitive, large telescope and show the pulsar pulses being detected in the side lobes as well as in the primary telescope beam.

\subsection{The astronomy signal processor}\label{sec:process}

The astronomy signal processor processes the incoming data streams and produces the final data products.  Typically this involves channelising the data streams. Dispersion smearing within a frequency channel is modelled by first simulating more channels than required and then averaging those channels to the requested output channelisation.  If a known pulsar is being observed then the data may first be coherently de-dispersed at the known DM of the pulsar and hence, in this case, there will be no channel-dependent dispersion smearing. 

The data volumes can be enormous and so the output data samples are typically written using only 1-, 2- or 8-bit quantisation.  Low-bit quantisation requires knowledge of the typical digital sample levels, which may change through an observation.    The levels can be pre-defined (e.g., any time sample above zero is set to 1 and any signal below to 0 for 1-bit data), or user-defined.  It is also possible to model automatic level-setting procedures. For instance, a running mean for each frequency channel (of specified number of samples) can be used to define the levels. This produces the observed change in the noise properties of the Lorimer burst detection after the event (in \FIG{fig:lorimer_burst}) and was used in the Parkes multibeam surveys that used an analogue filterbank system (e.g,. \citealt{Manchester+01MN}). The levels can also be set from the first samples and then held fixed for the remainder of the observation.  This is similar to the level setting used in the HTRU \citep{Keith+10MN} and Survey for Pulsars and Extragalactic Radio Bursts (SUPERB; \citealt{Keane+18MN}) at the Parkes radio telescope. 

Note that with low-bit digitisation it is common for a bright signal to saturate the system. This is shown in \FIG{fig:drift} where the pulses being detected by the simulated primary beam are so strong that the time series saturates.

\section{Injecting signals into actual observations}
\label{sec:inject}

Injecting simulated signals into actual pre-recorded data sets is often used to investigate the effectiveness of a given processing pipeline (e.g., \citealt{Gupta+21MN} and \citealt{Li+21Nat}). The challenge is that the recorded data are already quantised and so the injection must account for the expected survey sensitivity and the quantisation process. 

We assume Gaussian radiometer noise and that we know the frequency-dependent system temperature and telescope gain corresponding to the recorded data set. We then determine the probability that a source signal of specified amplitude will change the recorded bit.  For instance, as all signals have positive amplitude we note that a recorded 1 (in 1-bit data) will never become a 0.  However, there is a chance that a recorded 0 will become a 1.   In 2-bit data this becomes more challenging as we need to determine the probability that a recorded 0 remains as a 0 or becomes a 1, 2 or 3 (and similarly for other recorded values). The analytic results of these probability determinations are described in \APP{app:inject}.

\begin{figure}
    \centering
    \includegraphics[width=0.55\textwidth]{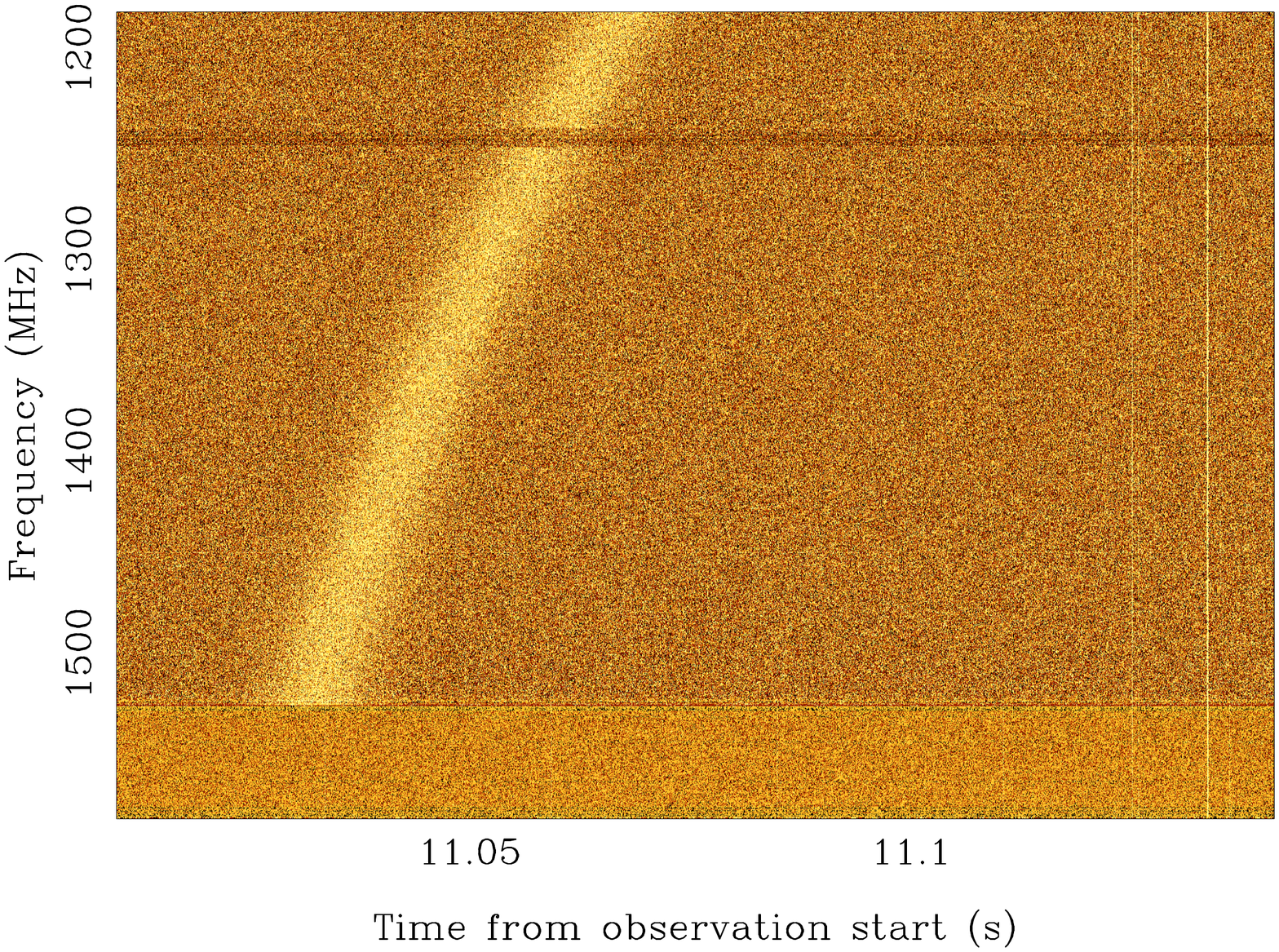}
    \caption{An injected single pulse event (DM = 30\,$\cmpc$) into real observations from the Parkes SUPERB survey (note that in this survey the observing frequency decreased with increasing channel number). The narrow-band and the impulsive RFI is real and present in the original data set.  The highest frequencies have significantly increased system temperature to model a hardware filter used to remove strong satellite signals. The \texttt{simulateBurst} and \texttt{injectSearchFile} utilities were used to implement the injection. The original data were obtained from file \textsc{bpsr141002\_144138\_beam01.sf} from \url{https://doi.org/10.4225/08/57EE7E6507372}.}
    \label{fig:inject}
\end{figure}

To demonstrate this method we inject a fake FRB in an archival data file from the SUPERB survey. The archived data file has been 2-bit sampled. We assume that the system temperature is 26\,K for the majority of the band, but increases significantly at the highest frequencies. This increase is caused by a filter that was used to mitigate the effect of satellites emitting in that part of the band (e.g., \citealt{Keane+18MN}).  We also note satellite interference around 1240\,MHz.  We also increase the expected system temperature in those bands.  We assume that the telescope gain, $G = 0.7$\,K/Jy.  In \FIG{fig:inject} we show the data set after the FRB (with a peak amplitude of 5\,Jy) has been injected.

\section{Discussion and conclusion}
\label{sec:disc}

We expect that there will be many uses of the simulation software described here. For instance, the software can be used to compare the effectiveness of algorithms developed to find specific astronomical signals. We may wish to know which is the most effective algorithm for detecting FRBs.  Such events are rare and so often such comparisons cannot be made using actual survey data sets.  Instead we can inject simulated FRBs into actual (or simulated) data sets. We then need to determine whether any signals identified by the algorithms were injected, or are false positives. The simulation software can add event labels into the output PSRFITS files. The event labels contain information such as the type, properties and time of each event. 

Owing to labelling, the simulator can be used to generate data for machine learning model training. Machine learning is increasingly used in radio astronomy to detect high-time resolution signals like pulsars \citep{Zhu+14ApJ, Morello+14MN} and FRBs \citep{Zhang+18ApJ, Zhang+20A&A}. Simulated data addresses data scarcity problems (such as accessibility and parameterisation) that occur when using real signals. The simulator also provides the means to represent the known properties of the signals, how different telescope observing systems can modify those signals (such as level setting procedures) and how those signals will appear in the local RFI environment.  Machine learning methods have a potential to learn such knowledge from the generated data and avoid learning spurious features from the real data. However, we note that training a machine learning model entirely on simulated data may lead to over-fitting as there always be differences between the simulated and the real data. 

For some algorithm comparisons it is necessary to restrict the number of events in given blocks of time.  For instance, the user may require that a given block of time has either 0 or 1 events and that no events should overlap (i.e., we should not have two pulses with different DMs overlapping each other).  Methods to constrain the event times in order to ensure this are available. For instance, the user can request that the FRB is injected at a random time that is constrained to be between two defined time intervals. The event label stored in the data file provides the exact time of the event.  Even though it is physically unrealistic we also provide an option that, if used, ensures level setting procedures (such as the recovery from a bright FRB event) does not affect any data in the adjacent block of data. 

This simulation code can also be used to benchmark pipelines for new surveys. For instance, the simulation code has been used to simulate an expected data set from the Parkes cryogenically cooled phased array feed. This involved simulating 76 beams and 2048 channels for each data stream with 2-bit digitisation. The output data volume was 646\,GB of data in total for an observation of 1000 seconds.  We made use of these data streams to benchmark different processing algorithms and to confirm that the infrastructure was in place to record, transfer, process and archive such massive data volumes.  The simulation software has been divided to allow easy parallelisation of tasks.  For instance, different processors can process different telescope beams, or one processor could simulate radiometer noise, whereas another processor simulates expected FRB events.

An earlier version of this simulation software was used by \cite{Li+21Nat} who injected fast radio burst signals into FAST observations in order to determine the sensitivity of their survey and the completeness factor for their pipeline.  The simulated PSRFITS files are in the same format as the data from the telescope and so any pipeline that has been developed for actual data sets can easily be run and tested on the simulated data sets.

It is impossible to predict every possible signal that may be detectable and some source types (RFI in particular) are complex and more detailed simulations could be developed. The software is developed so that new simulated signals can easily be produced.  We will also continue to develop the software that produces the output data products. In the future it is likely that more use will be made of calibrated data streams with polarisation information. Currently we assume the data sets represent the total intensity (Stokes I), but plan to update the code to enable the simulation of all four Stokes parameters. 

How realistic could we make the simulations? For a given, impulsive event observed in real data (such as a FRB) it is likely that a realistic simulation of that event (noting its frequency and time structure) and the noise properties of the underlying noise could be made.  The primary challenge is in modelling the longer-term system variations, such as changes in the background noise caused by spill-over, long-term instrumental gain variations, or structural deformation of the telescope at different pointing directions. A simulation that includes a detailed model of the telescope structure and its surroundings would require that the electromagnetic waves and measured voltages (for two polarisation channels) are simulated, in contrast to the simulations here of detected, channelised and quantised, single polarisation data streams. However, even with the existing simulation and the ability to inject simulated data sets in existing data sets, thought must be given to malicious use whereby a user injects a signal of interest into an actual observation and then claims it to be a real detection. A similar scenario would be an injection made for non-malicious purposes, but a subsequent user obtaining that data set without knowing an injection had been made.

The LIGO/Virgo Collaboration  use blind injections of fake gravitational wave (GW) signals to test the data analyses from multiple independent working groups. All the GW detections are verified after comparison with blind injected signals, including the first black hole-black hole event GW~150924 \citep{LSC16PhRvL}.  However, after their first discovery it was essential for the LIGO/Virgo team to ensure that their signal was not a malicious injection. Radio astronomy archives, such as a Parkes-telescope pulsar archive \citep{Hobbs+11PASA}, record a check-sum along with each observation to ensure that any modification of the raw data after archiving can easily be identified. 

In this paper we have described the software package. We are now using this software to make a data challenge that will contain injected signals into both real and simulated data sets. We will use those data to develop algorithms that can be used both to find the ``known unknowns'' such as pulsars and FRBs, but also to find the ``unknown unknowns'' in our massive data volumes.


\section*{Acknowledgements}

We make use of archival data obtained from the Parkes radio telescope. The Parkes radio telescope is part of the Australia Telescope National Facility (\url{https://ror.org/05qajvd42}) which is funded by the Australian Government for operation as a National Facility managed by CSIRO.  We acknowledge the Wiradjuri people as the traditional owners of the Observatory site.  This paper includes archived data obtained through the Parkes Pulsar Data archive on the CSIRO Data Access Portal (http://data.csiro.au). We thank Phil Edwards for reading and commenting on the manuscript. We also particularly acknowledge Emil Lenc for providing us the 1-bit data of the cat image ($96\times96$ pixels) used in this paper.

\section*{Data availability}

All the data sets described in this paper were either simulated through the simulateSearch software package that is available from \url{https://bitbucket.csiro.au/scm/psrsoft/simulatesearch.git} or from publicly available Parkes data sets that can be downloaded from \url{https://data.csiro.au}. Links to the exact observations used here are provided in the Figure captions.



\bibliographystyle{mnras}
\bibliography{refs} 




\appendix

\section{The binary data file format}
\label{app:format}

The simulation code produces a set of binary files representing the signals being simulated. Those files then get combined and quantised into the final PSRFITS-format file.

The binary files contain header information followed by the raw data stream, which represents each frequency channel for each time sample as a 32-bit floating point value.  The header is stored as follows:
\begin{itemize}
    \item format number stored in 64 characters. The current header versions are ``FORMAT 1'', ``FORMAT 1.1'', ``FORMAT 1.2'' and ``FORMAT 2.1''.  Here we describe FORMAT 1.2.
    \item data set name (128 characters)
    \item Start time (t0; seconds) (32-bit float)
    \item End time (t1; seconds) (32-bit float)
    \item Sampling time (seconds) (32-bit float)
    \item Frequency of first channel (MHz) (32-bit float)
    \item Frequency of last channel (MHz) (32-bit float)
    \item Number of channels (integer)
    \item Positional information type (integer)
\end{itemize}
If the positional information type is 1 then we next record:
\begin{itemize}
    \item Source right ascension (radians) (32-bit float)
    \item Source declination (radians) (32-bit float)
\end{itemize}
otherwise a filename containing positional information is provided as 128 characters.  The header then contains:
\begin{itemize}
    \item Flag indicated use of angles (1 or 0) as single byte character
    \item Initial random number seed (long integer)
    \item Flag indicating the existence of labels (1 or 0) as integer
\end{itemize}
If event labels are present in the data set then the header information contains the number of event labels (long integer) and then for each event:
\begin{itemize}
 \item Type of event (32 characters)
 \item Properties of event (128 characters)
 \item Frequency frequency for time (MHz) (32-bit float)
 \item Flag indicated how the time is stored (integer)
 \item Dispersion measure of event (32-bit float)
 \item Start time of event (seconds) (32-bit float)
 \item End time of event (seconds) (32-bit float)
 \item Flag indicating how the frequency information is stored (integer)
 \item Initial frequency (MHz) of event (32-bit float)
 \item Final frequency (MHz) of event (32-bit float)
 \item Amplitude of event (32-bit float)
 \end{itemize}

It is possible to write out individual samples for the representation of radiometer noise, but such a data set cannot be compressed (as it consists of noise), but can simply be described by a small number of Gaussian amplitude values.  We therefore provide the ability to write out the Gaussian amplitudes for each channel as a binary 32-bit floating point value.

For rare events we also do not need to write out a representation of the signal across the entire data span (as it may only occur for a millisecond or so in a many hour observation).  A compressed binary data file consists of:
\begin{itemize}
    \item Number of events (integer)
\end{itemize}
and then for each event:
\begin{itemize}
    \item Number of samples for this event (integer)
    \item Number of frequency channels (integer)
    \item Start time of event (32-bit float)
    \item Values for each sample in this event (Nsamples x Nchannels 32-bit floats)
\end{itemize}

We provide an example on how to build the binary file using Python, which can be found in the tutorial of this software in the BitBucket repository \url{https://bitbucket.csiro.au/scm/psrsoft/simulatesearch.git}.

\section{Injecting into quantised data streams}
\label{app:inject}

We assume that the background noise (which has already been quantised) represents a Gaussian distribution. The probability density function of the noise signal, $\Noi$, is given as
\begin{equation}
    P(\Noi)=\frac{1}{\sigma\sqrt{2\pi}}e^{-\frac{(\Noi-\mu)^2}{2\sigma^2}}.
\end{equation}

The noise level $\sigma$ is described by the Radiometer Equation
\begin{equation}
    \sigma=\frac{T_\mathrm{sys}}{G\sqrt{N_\mathrm{p} t_\mathrm{samp} \Delta\nu}},
\end{equation}
where $G$ is the telescope gain in units of $\rm K/Jy$, $T_\mathrm{sys}$ is the system temperature, $t_\mathrm{samp}$ is the digital sampling time and $\Delta\nu$ is the receiver bandwidth.

The cumulative distribution function of noise is given by
\begin{equation}
    \Phi(\Noi)=\frac{1}{2}\left[1+\mathrm{erf}\left(\frac{\Noi-\mu}{\sigma\sqrt{2}}\right)\right],
\end{equation}

For 1-bit case, there is only one threshold to change the digitised signal, i.e., the mean $\mu$.  For a signal $\Sig$ injected to data samples, we have the digitised values as follow
\begin{equation}
    X(\Sig+\Noi)=
    \begin{cases}
    1 \hspace{6em} \Sig+\Noi\ge\mu\\
    0 \hspace{6em} \Sig+\Noi<\mu\\
    \end{cases}.
\end{equation}

The probability of four cases in digit changes can be obtained as
\begin{equation}
    P_{ij} = \begin{pmatrix}
    \frac{\Phi(\mu-\Sig)}{0.5} &1- \frac{\Phi(\mu-\Sig)}{0.5} \\
    0 & 1 \\
    \end{pmatrix},
\end{equation}
where $i$ and $j$ are 0 or 1.

For injection into 2-bit data we need to determine the 10 probabilities. Here we present the probability for 2-bit case, the analogous signal intensity can be digitised as 0, 1, 2, 3. The corresponding thresholds are set as:
\begin{equation}
    X(\Sig+\Noi)=
    \begin{cases}
    0 \hspace{6em} \Sig+\Noi<\mu-l\sigma \\
    1 \hspace{6em} \mu-l\sigma\le\Sig+\Noi<\mu \\
    2 \hspace{6em} \mu\le\Sig+\Noi<\mu+l\sigma \\
    3 \hspace{6em} \Sig+\Noi\ge\mu+l\sigma \\
    \end{cases},
\end{equation}
where $l$ is the level setting number for digitisation. For 2-bit data, we usually adopt $l=0.9674$ \citep{JA98PASP, KV01A&A}. 

There are 16 cases for the digitisation by injected signal, which can be described as the following 4$\times$4 matrix
\begin{equation}
    P_{ij} = \begin{pmatrix}
    P_{00} & P_{01} & P_{02} & P_{03} \\
    0 & P_{11} & P_{12} & P_{13} \\
    0 & 0 & P_{22} & P_{23} \\
    0 & 0 & 0 & P_{33} \\
    \end{pmatrix},
\end{equation}
where
\begin{equation}
    \begin{split}
        P_{00}&=\frac{\Phi(\mu-l\sigma-\Sig)}{\Phi(\mu-l\sigma)}, \\
        P_{01}&=\frac{\Phi\left[\min(\mu-\Sig, \mu-l\sigma)\right]-\Phi(\mu-l\sigma-\Sig)}{\Phi(\mu-l\sigma)}, \\
        P_{02}&=\frac{\Phi\left[\min(\mu+l\sigma-\Sig, \mu-l\sigma)\right]-\Phi(\mu-\Sig)}{\Phi(\mu-l\sigma)}, \\
        P_{03}&=\frac{\Phi(\mu-l\sigma)-\Phi(\mu+l\sigma-\Sig)}{\Phi(\mu-l\sigma)}, \\
        P_{11}&=\frac{\Phi(\mu-\Sig)-\Phi(\mu-l\sigma)}{0.5-\Phi(\mu-l\sigma)}, \\
        P_{12}&=\frac{\Phi\left[\min(\mu+l\sigma-\Sig, \mu)\right]-\Phi\left[\max(\mu-l\sigma, \mu-\Sig)\right]}{0.5-\Phi(\mu-l\sigma)}, \\
        P_{13}&=\frac{0.5-\Phi\left[\max(\mu+l\sigma-\Sig, \mu-l\sigma)\right]}{0.5-\Phi(\mu-l\sigma)}, \\
        P_{22}&=\frac{\Phi(\mu+l\sigma-\Sig)-0.5}{\Phi(\mu+l\sigma)-0.5}, \\
        P_{23}&=\frac{\Phi(\mu+l\sigma)-\Phi\left[\max(\mu-l\sigma-\Sig, \mu)\right]}{\Phi(\mu+l\sigma)-0.5}, \\
        P_{33}&=\frac{1}{1-\Phi(\mu+l\sigma)}.
    \end{split}
\end{equation}

\section{Making the images in this paper}\label{sec:makeImages}

To reproduce the figures in this paper, we provide parameter files and the commands used. For \FIG{fig:frb}. We create a telescope data file, whose system has 96 frequency channels (pks\_96chan.params):
\begin{verbatim}
name: System for a mock radio telescope
observer: rluo
f1: 1230
f2: 1518
nchan: 96
nsblk: 2048
t0: 0
t1: 10
tsamp: 256e-6
raj: 0
decj: 0
useAngle: 0
gain: 0.7
tsys: 25
nbits: 1
imjd: 58456
smjd: 36400
\end{verbatim}
To make Figure 1 we required the parameters for the FRB (stored in frb.params):
\begin{verbatim}
dmburst: 4.9 1400 1 -2 0.005 300 2
\end{verbatim}
and ran the following commands to make Figure 1:
\begin{verbatim}
simulateSystemNoise -p pks_96chan.params -o noise.dat
simulateBurst -p pks_96chan.params -p frb.params -o frb.dat
createSearchFile -p pks_96chan.params -f noise.dat -f frb.dat -o frb.sf
pfits_plot -f frb.sf -s1 9 -s2 9
\end{verbatim}

For the other ten figures in this paper, we wrap the used parameter files and simulation commands into the tutorials of this software, which can also be found in the repository \url{https://bitbucket.csiro.au/scm/psrsoft/simulatesearch.git}

\bsp	

\label{lastpage}
\end{document}